\documentclass[sigconf]{acmart}

\usepackage{libertine}
\usepackage{xspace}
\usepackage{amsmath}
\usepackage{amssymb}  
\usepackage{balance}
\usepackage{xspace}

\usepackage{mathtools}
\usepackage{graphicx}
\usepackage{algorithm,algorithmicx,algpseudocode}  

\newcommand{\neatfa}{\textsc{NeatFA}\xspace}

\title{On the Minimal Set of Inputs Required for Efficient  Neuro-Evolved Foraging}

\author{John Ericksen}
\affiliation{Moses Biological Computation Lab}
\affiliation{University of New Mexico, Albuquerque, USA.}
\email{johncarl@unm.edu}

\author{Abhinav Aggarwal}
\affiliation{Department of Computer Science}
\affiliation{University of New Mexico, Albuquerque, USA.}
\email{abhiag@unm.edu}

\author{Melanie E. Moses}
\affiliation{Moses Biological Computation Lab}
\affiliation{University of New Mexico, Albuquerque, USA.}
\email{melaniem@unm.edu}

\setcopyright{acmcopyright} % if you give the rights to ACM
\acmDOI{...} % DOI - Insert your DOI below...
\acmISBN{...} % ISBN - Insert your conference/workshop's ISBN below...
\acmYear{2019} % Insert Publication year
\copyrightyear{2019} % Insert Copyright year (typically the same as above)
\acmConference[BDA'19]{The 7th Workshop on Biological Distributed Algorithms (BDA 2019)}{July 29, 2019}{Toronto, CA.}



\begin{document}
\begin{abstract}
In this paper, we perform an ablation study of \neatfa, a neuro-evolved foraging algorithm that has recently been shown to forage efficiently under different resource distributions. Through selective disabling of input signals, we identify a \emph{sufficiently} minimal set of input features that contribute the most towards determining search trajectories which favor high resource collection rates. Our experiments reveal that, independent of how the resources are distributed in the arena, the signals involved in imparting the controller the ability to switch from searching of resources to transporting them back to the nest are the most critical. Additionally, we find that pheromones play a key role in boosting performance of the controller by providing signals for informed locomotion in search for unforaged resources. 
\end{abstract}

\maketitle

\section{Introduction}

The foraging problem is a well studied challenge in swarm robotics.  For a robot swarm to successfully forage for resources, individual robots in the swarm must collectively solve a series of sub-tasks~\cite{liu2007towards, winfield2009foraging}.  First, the robots must leave the base station (nest) and enter a search phase to canvas the environment for resources.  Once a resource is identified, the robots store the resource for transport and enter the return-to-nest phase.  In this phase, the robots must find a path back to the nest using some environmental clue or knowledge of their surroundings.  Finally, upon returning to the nest, the must robots deliver the collected resource, emptying their resource hold, and return to the search phase.  This cycle is repeated across all robots in the swarm until some condition is met - either enough resources have been collected, or resources have been exhausted in a region for example.

Central to solving the foraging problem is the design of the robot controllers in the swarm.  Each controller is responsible for directing robots through the phases of foraging and perform any cursory tasks that may be useful to the swarm (i.e. laying pheromones).  In previous work \cite{ericksen2017automatically}, we preset a neural network controller built using Neuroevolution of Augmenting Topologies (NEAT) evolved to solve the foraging problem.  We call this controller the Neat Foraging Algorithm, or \neatfa.  This controller is shown to perform well in comparison to other algorithms, specifically the hand-build Distributed Deterministic Spiral Algorithm (DDSA) \cite{fricke2016distributed} and the genetically tuned Central Place Foraging Algorithm (CPFA) \cite{hecker2015beyond}.  Included in this study is using NEAT to evolve resource distribution (clustered, semi-clustered, and uniformly randomized) specific controllers for comparison and a resource-agnostic "general" controller evolved against all distributions.

Despite the positive results from \neatfa, the blackbox nature of the evolved neural network controller left insights about the behavior lacking. In some preliminary work \cite{ericksen2017automatically}, it was hypothesized that certain inputs, like pheromones, were used and critical to the behavior of \neatfa. However, no formal analysis or empirical arguments were provided in support of this. 

In this work, we present a blackbox analysis of the controllers evolved for specific resource distributions.  We analyze the controllers by removing groups of inputs (known as ablation in the medical field) to the network and observing how the controller behavior changes. More specifically, we focus on how the foraging efficiency (the number of resources collected with time) gets affected when one or more inputs to the network are disabled. We do not, however, re-evolve the controllers to accommodate these changes and simply use the same network structure with the inputs disabled.  

With this strategy we answer two key questions.  First, \emph{which inputs may be removed without drastically affecting the performance or behavior of the controller?}  This helps advise which inputs are necessary for controller to function and which inputs an evolved approach decided to leverage. More importantly, it helps identify signals which potentially lead to the decisions taken by the controller to help collect the resources faster. Second, after the key inputs are identified, we determine \emph{which inputs drive the phases of foraging?} In other words, what inputs play a key role in helping the controller determine when to switch from one phase of the search to another. This is critical to the observed behavior of the evolved algorithm since NEAT is designed to favor those strategies that not only find resources, but also bring them back to the nest (see Section~\ref{sec:neatfa} for details). Intuition tells us that inputs like holding-food drives the transition between the search phase and the return-to-nest phase, and nest-sight guides the robots back to the nest. Our ablation study verifies this intuitive claim in addition to suggesting certain other important inputs for particular resource distributions.

The rest of the paper is organized as follows. We first provide a brief description of \neatfa and then discuss the different experiments in our ablation study. Next, we discuss the different results obtained and infer a sufficiently minimal set of inputs required for the \neatfa to forage efficiently.

\section{The \neatfa controller}
\label{sec:neatfa}

The \neatfa controller is built using connected weighted perceptrons~\cite{mitchell1997artificial} with a sigmoid logistic function\footnote{For flexibility, the neural network is not restricted to a feed-forward structure, rather perceptrons in the network may connect to any other perceptron making the network potentially cyclic. In addition, the network is updated using a sampling strategy to avoid issues that arise from network cycles.}. Inputs and outputs to/from the \neatfa network are chosen to mirror the DDSA and CPFA inputs and outputs, and meant to support an implementation of the foraging problem (see~\cite{ericksen2017automatically} for details).  Inputs to the network are sensor readings that provide the following signals:
\begin{enumerate}  
\item \emph{Compass Inputs}: Given as real values X, Y, Z, and W from a Quaternion that specifies the robot's orientation in the ARGoS simulation~\cite{argos}. Each of these 4 real values are clamped on to the network at separate input neurons.
  \item \emph{Detection of Holding a Resource}: A single Boolean value which is set when the robot is holding a resource. 
  \item \emph{Detection of Proximity to a Resource}: A single Boolean value which is set when the robot is within the (pre-specified) collection radius of a food resource.  As a caveat, note that this signal triggers only when the robot is already carrying another food item and it happens to be in the detection proximity of a resource.
  \item \emph{Sight to the Nest}: A set of four real valued cardinal inputs, each of which is the maximum of 6 light magnitude inputs from each of the 4 cardinal sides (top, left, bottom and right).
  \item \emph{Detection of Proximity to a Pheromone}: A single Boolean value which is set when the robot is within the (pre-specified) collection radius of an already laid pheromone. This pheromone could have been laid by the same robot or some other robot in the arena. The controller is unable to distinguish the identity of the robot that laid the pheromone it just detected. 
  \item \emph{Detection of Proximity to Other Robots}: A set of four real value cardinal inputs, each of which is the maximum of 6 distance inputs from each of the 4 cardinal sides (similar to that for the nest sight).
\end{enumerate}

\noindent Outputs from the \neatfa network are as follows:

\begin{enumerate}
  \item \emph{Left and Right Wheel Speeds}: These are sampled from two output neurons, scaled to match the minimum (negative) and maximum wheel speeds (-16 and 16 units respectively) and set to the linear velocity method on the given robot wheel motors.
  \item \emph{Lay Pheromone}: This is sampled from one output neuron, a positive signal from which directs the robot to lay a pheromone trail at its current location.
\end{enumerate}

To drive evolution with NEAT, the following fitness function is used to mirror the foraging phases: one point per resource collected and two points for dropping collected resources at the nest. Furthermore, to capture the effect of different resource distributions in the arena, NEAT was used to generate a controller for each of the 3 commonly studied resource distributions: clustered (where resources are present in clusters of equal size, distributed uniformly at random across the arena), semi-clustered (where resources are distributed in clusters of different sizes, typically driven by a power law distribution governing the number of clusters of a given size), and uniform (where resources are placed uniformly at random inside the arena).  Each of these controllers exhibit similar, but not identical foraging behaviors. 

\section{Our Experiments}
To identify the most important input features that NEAT used to evolve the foraging pattern observed in \neatfa, we perform an ablation study in which we disable individual inputs, one feature at a time, and observe the change in the average number of resources collected over time. To disable any given feature, we introduced a static Boolean value for the corresponding input to indicate that it is turned off. We keep the random seed identical between different simulation trials. This allows us to keep simulation environments (i.e. resource distribution) identical across different experiments and isolates the behavior differences between disabled inputs.

In this paper, we perform our ablation study on a single robot system. This helps us focus on the foraging pattern that a single robot uses. Needless to say, since there are no additional robots in the arena, disabling the sensor that determines proximity to other robots has minimal effect on the foraging efficiency, as is evident in the results we obtain. Note that reducing the swarm size to one eliminates any interaction or nest congestion that may be mitigated by a particular input feature.  Also, this reduces the effect pheromones could have for communicating information within the swarm. We do not currently study the effect of inter-robot communication in this paper.

\vspace{0.8em}
\noindent  \textbf{Methodology.} We seek a sufficiently minimal set of inputs that drives efficient foraging of resources in the arena. By sufficient, we mean that the foraging efficiency when only these minimal inputs are enabled is reasonably high. By minimality, we mean that any fewer inputs will cause a drastic decrease in the foraging efficiency. Identifying such a set helps advise not only our experiments, but other potential foraging algorithms to what are some key factors that determine the optimal trajectory for resource collection. 

To do this, we execute a baseline controller and compare the number of seeds collected over a simulation run to controllers with each features disabled.  This comparison is plotted and the features with little to no impact on the controller's resource collection count over time.

Following this first round of analysis, we assemble the sufficiently minimal set of inputs for each controller by distribution.  To establish minimality, we remove one feature from the set and count the number of resources collected. Additionally, we enable one extra feature to the controller and count the number of resources now collected. If the former shows a significant drop in resources collected and the latter show no significant change in resources collected, we posit that we have a (sufficiently) minimal set of inputs for the given controller.

\section{Overview of our Results}
We find that while NEAT leverages a majority of the inputs for each controller, however, some inputs affect foraging more significantly than others when removed. 

\begin{figure}[t]
    \centering
    \includegraphics[width=\columnwidth]{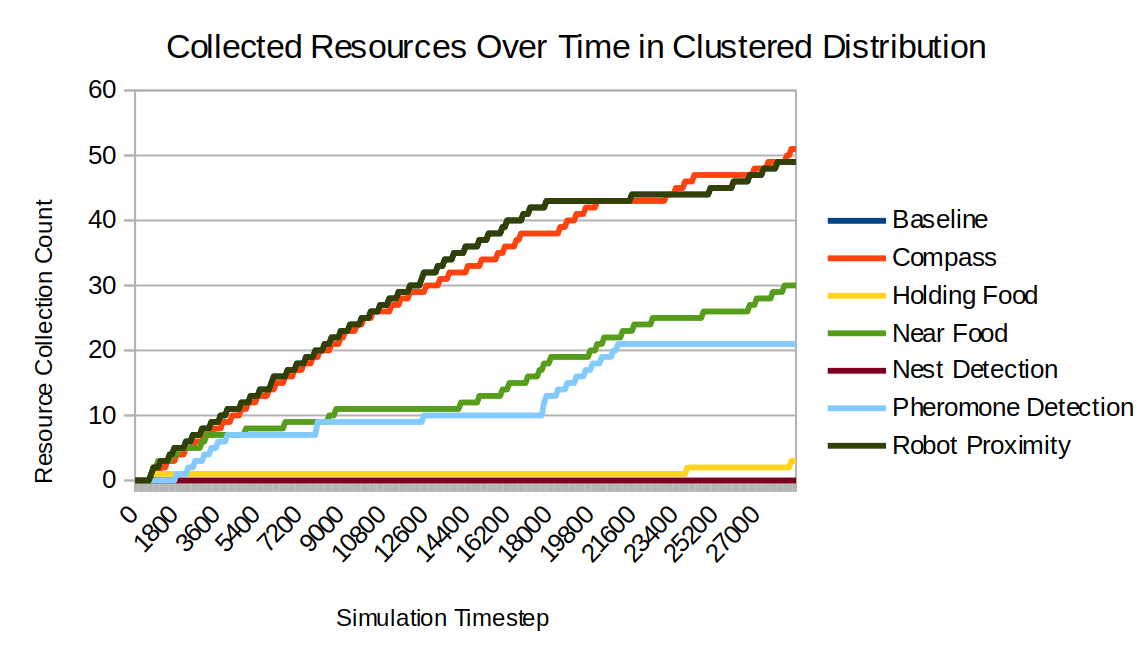}
    \caption{Plots of the ablation study for an arena with clustered resources.\vspace{-1em}}
    \label{fig:clustered_minimal_set}
\end{figure}

\begin{figure}[t]
    \centering
    \includegraphics[width=\columnwidth]{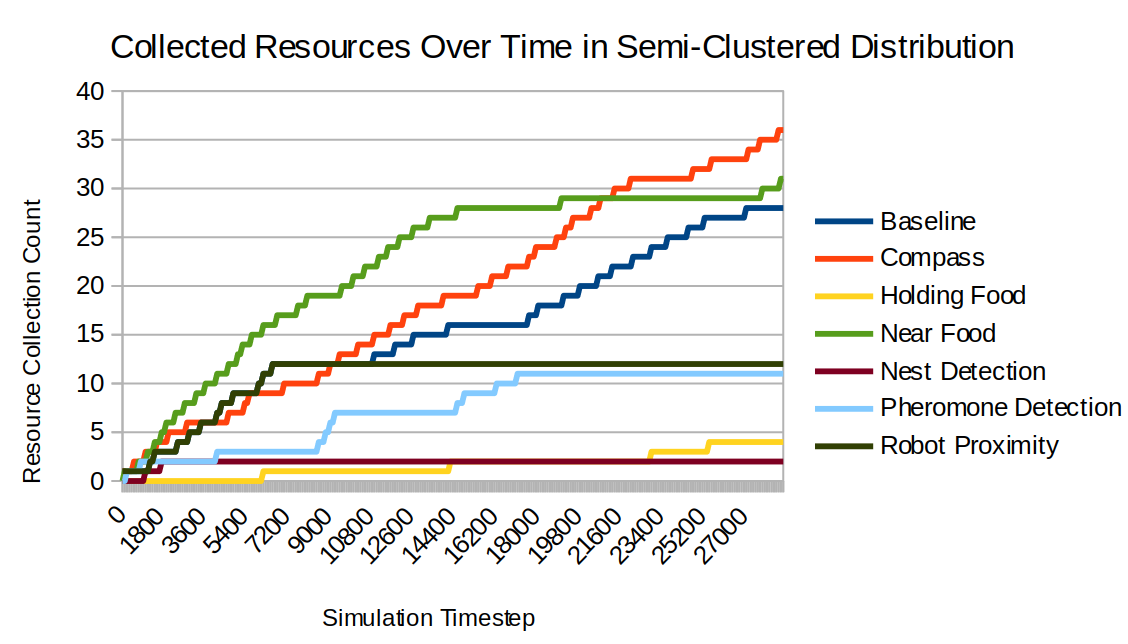}
    \caption{Plots of the ablation study for an arena with semi-clustered resources.\vspace{-1em}}
    \label{fig:semiclustered_minimal_set}
\end{figure}

\begin{figure}[t]
    \centering
    \includegraphics[width=\columnwidth]{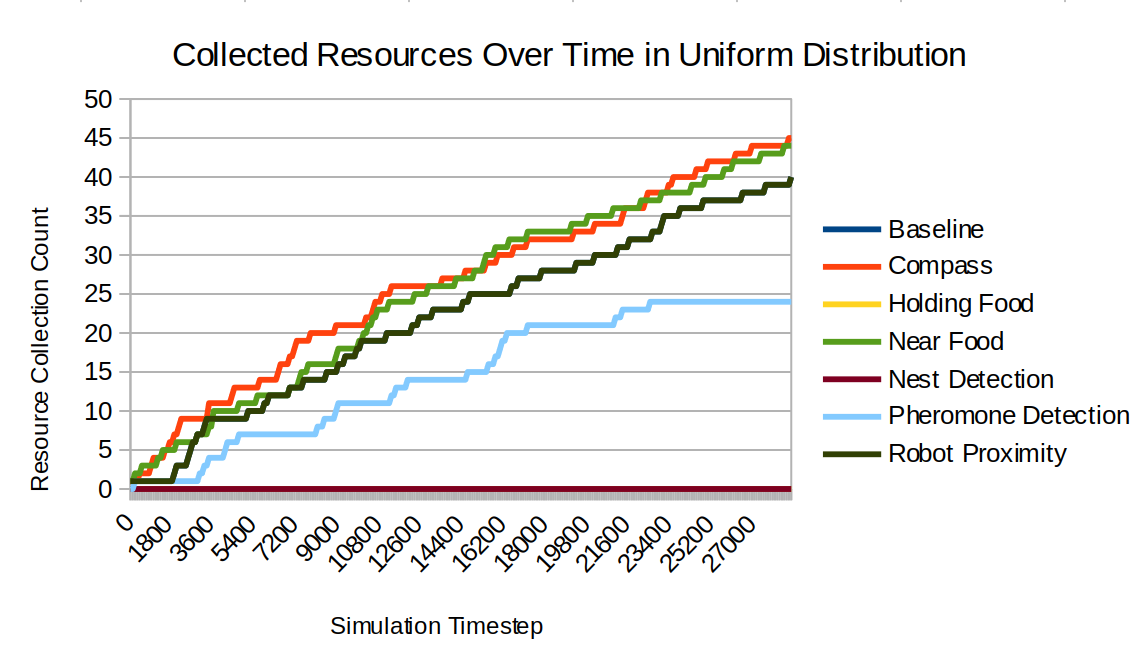}
    \caption{Plots of the ablation study for an arena with uniformly distributed resources.\vspace{-1em}}
    \label{fig:uniform_minimal_set}
\end{figure}

Across all distributions, we find that the Compass input makes no negative impact on the controller performance when removed. For the clustered distribution (see Figure \ref{fig:clustered_minimal_set}), the Robot Proximity makes no significant impact on performance.  For the semi-clustered distribution (see Figure \ref{fig:semiclustered_minimal_set}), The Near Food input makes no negative impact on performance.  Finally, for the uniform distribution (see Figure \ref{fig:uniform_minimal_set}), the Neat food and Robot Proximity inputs have little significant impact on performance when removed. For each distribution, we also find that the Nest Detection and Holding Food features, when removed, reduce the number of resources collected to near zero.  Interestingly, the Pheromone input, when disabled, does affect performance, generally cutting the number of collected resources in half.

\begin{figure}[t]
    \centering
    \includegraphics[width=\columnwidth]{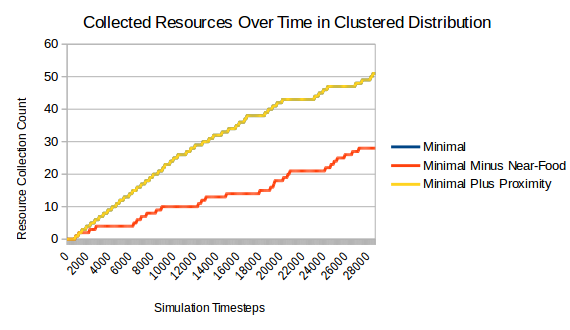}
    \caption{Plots of foraging efficiency to study the sufficiently minimal set of inputs required for the controller evolved for the clustered distribution.\vspace{-1em}}
    \label{fig:clustered_minimal_combined}
\end{figure}

For the clustered distribution, we identify that Holding Food, Nest Detection, Near Food, Pheromones, and Robot Proximity inputs as sufficiently minimal (see Figure \ref{fig:clustered_minimal_combined}).  These inputs produce a count of 51 resources.  Adding Robot Proximity back into the controller produces a count of 51 resources and removing the Near Food feature cuts the performance down to 28 resources collected.  

\begin{figure}[t]
    \centering
    \includegraphics[width=\columnwidth]{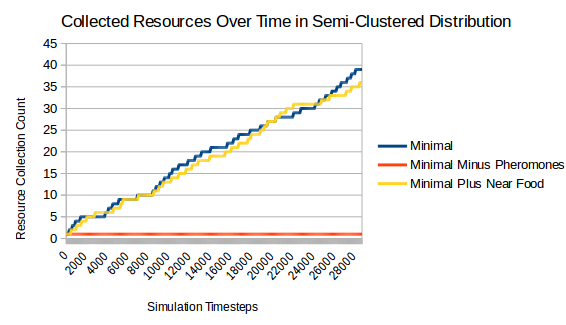}
    \caption{Plots of foraging efficiency to study the sufficiently minimal set of inputs required for the controller evolved for the semi-clustered distribution.\vspace{-1em}}
    \label{fig:semiclustered_minimal_combined}
\end{figure}

For the semi-clustered distribution controller we identify the Holding Food, Nest Detection, Pheromones, and Near Food inputs as sufficiently minimal (see Figure \ref{fig:semiclustered_minimal_combined}).  These inputs produce a count of 39 resources.  Adding Near Food Proximity back into the controller produces a count of 36 resources and removing the Pheromones feature cuts the performance down to 1 resource collected. 

\begin{figure}[t]
    \centering
    \includegraphics[width=\columnwidth]{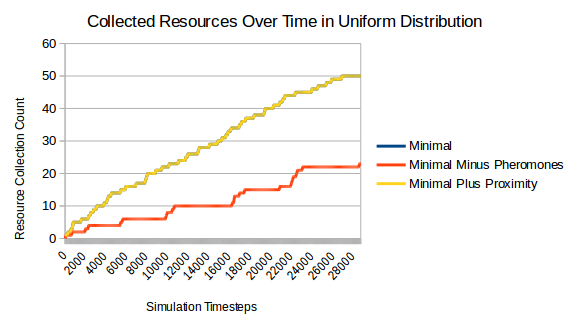}
    \caption{Plots of foraging efficiency to study the sufficiently minimal set of inputs required for the controller evolved for the uniformly random distribution.\vspace{-1em}}
    \label{fig:uniform_minimal_combined}
\end{figure}

Finally, for the uniform distribution controller we identify the Holding Food, Nest Detection, and Pheromones inputs as sufficiently minimal (see Figure \ref{fig:uniform_minimal_combined}).  These inputs produce a count of 50 resources.  Adding Near Food Proximity back into the controller produces a count of 50 resources and removing the Pheromones feature cuts the performance down to 23 resource collected.

\section{Discussion}
To analyze the behavior of the sufficiently minimal controllers, we focus on the input features with the most impact across all distributions: the Hold Food detection, Nest Light detection, and the Pheromone laying and detection features.

\subsection{Detection of Holding Food}
The ability to hold targets provides two important functionalities to the controller: (1) collection of different resources from the arena and transporting them back to the nest; and (2) switch from search phase to the transport phase. The second functionality is extremely critical as it allows the controller to start the reverse process of finding its way back to the nest. 

As observed, switching off the ability to hold targets has most significant effect on this transition. As expected, the controller is no longer able to find its way back to the nest. However, this inability is not due to any navigational handicap, but merely due to the fact that the controller does not know when to start searching for a path back to the nest. Since it has no way of telling whether it has picked up a resource and the only trigger for switching from resource-search to nest-search is holding a target in hand, the controller always believes that it is searching for resources in the arena. 

\subsection{Detection of Nest Light}
Detection of the nest light signal is intended to give the controller a mechanism to find a path back to the nest.  This allows the controller to orient itself to the nest and make movements towards it.  This is a critical feature for the controller during the transport phase once a resource is collected with the next step being to deliver this resource to the nest.

In all three seed distributions, turning off the nest light signal had the same outcome.  The controller was unable to orient itself to the nest for a direct path back to deliver a seed. Without this orientation, each controller searches aimlessly to find the nest, only occasionally delivering a resource to the nest in lucky circumstances.  In fact, for the clustered environment, the absence of a nest light signal resulted in zero resource collected in a full trial run of the simulation.

\subsection{Ability to Lay and Detect Pheromones}

Pheromone laying and detection is intended to give the robot a mechanism to store and communicate location information throughout the swarm, including itself. For the single robot case, however, \neatfa seems to use pheromones as a mechanism to draw a region that the swarm is working within. This is manifested by robots turning sharper back into the pheromone trails after no pheromone is detected. Disabling pheromone has an interesting effect -- the robots hesitate to move forward. This is a strong evidence towards the fact that pheromones are used in some regard for locomotion in the environment. After setting the input signal to a constant 1 value, across all three distributions the performance was not completely restricted, but showed a drop in collected resources of one half.  Therefore, pheromones are not critical to the controller's behavior, but they are a significant factor in higher performance.

\subsection{Conclusions and Future Work}
Our ablation study on \neatfa determined the ability to detect holding a resource and finding the way back to the nest by following the nest light as two most important signals that drive the efficiency of this algorithm. While some other input signals become more important in certain environments, our experiments show that laying pheromones plays an important role in deciding the search trajectory of the robot. An interesting open question is to study the effect of cross communication between different robots through these pheromone trails on the collective foraging efficiency of the swarm. While site fidelity is certainly an observation made for search patterns on a clustered environment, it is also interesting to ask what signals cause the robots to explore areas that have not been previously explored.

\bibliography{ref.bib}

\begin{thebibliography}{1}

\bibitem{argos}
Argos.
\newblock \url{http://www.argos-sim.info/}.
\newblock Retrieved 2016-09-06.

\bibitem{ericksen2017automatically}
{\sc Ericksen, J., Moses, M., and Forrest, S.}
\newblock Automatically evolving a general controller for robot swarms.
\newblock In {\em 2017 IEEE Symposium Series on Computational Intelligence
  (SSCI)\/} (2017), IEEE, pp.~1--8.

\bibitem{fricke2016distributed}
{\sc Fricke, G.~M., Hecker, J.~P., Griego, A.~D., Tran, L.~T., and Moses,
  M.~E.}
\newblock A distributed deterministic spiral search algorithm for swarms.
\newblock In {\em Intelligent Robots and Systems (IROS), 2016 IEEE/RSJ
  International Conference on\/} (2016), IEEE, pp.~4430--4436.

\bibitem{hecker2015beyond}
{\sc Hecker, J.~P., and Moses, M.~E.}
\newblock Beyond pheromones: evolving error-tolerant, flexible, and scalable
  ant-inspired robot swarms.
\newblock {\em Swarm Intelligence 9}, 1 (2015), 43--70.

\bibitem{liu2007towards}
{\sc Liu, W., Winfield, A.~F., Sa, J., Chen, J., and Dou, L.}
\newblock Towards energy optimization: Emergent task allocation in a swarm of
  foraging robots.
\newblock {\em Adaptive behavior 15}, 3 (2007), 289--305.

\bibitem{mitchell1997artificial}
{\sc Mitchell, T.~M.}
\newblock Artificial neural networks.
\newblock {\em Machine learning 45\/} (1997), 81--127.

\bibitem{winfield2009foraging}
{\sc Winfield, A.~F.}
\newblock Foraging robots.
\newblock {\em Encyclopedia of complexity and systems science\/} (2009),
  3682--3700.

\end{thebibliography}
\bibliographystyle{acm}
\end{document}